\documentclass[conference]{IEEEtran}
\IEEEoverridecommandlockouts
\usepackage{cite}
\usepackage{amsmath,amssymb,amsfonts}
\usepackage{algorithmic}
\usepackage{graphicx}
\usepackage{textcomp}
\usepackage{xcolor}
\def\BibTeX{{\rm B\kern-.05em{\sc i\kern-.025em b}\kern-.08em
    T\kern-.1667em\lower.7ex\hbox{E}\kern-.125emX}}

\usepackage{comment}

\begin{document}

\title{Brain tumor grade classification Using LSTM Neural Networks with Domain Pre-Transforms}

\author{\IEEEauthorblockN{Maedeh Sadat Fasihi and Wasfy B. Mikhael, Fellow, IEEE}
\IEEEauthorblockA{Department of Electrical Engineering and Computer Science,
University of Central Florida,
Orlando, FL, 32816\\
Emails: fasihi\_m@knights.ucf.edu, Wasfy.Mikhael@ucf.edu}
}
\maketitle
\begin{abstract}
The performance of image classification
methods heavily relies on the high-quality annotations, which are not
easily affordable, particularly for medical data. To alleviate this limitation, in this study, we propose a weakly supervised image classification
method based on combination of hand-crafted features.
We hypothesize that integration of these hand-crafted features alongside  Long short-term memory (LSTM) classifier can reduce the adverse effects of weak labels in classification accuracy. Our proposed algorithm is based on selecting the appropriate domain representations of the data in Wavelet and Discrete Cosine Transform (DCT) domains. This information is then fed into LSTM network to account for the sequential nature of the data. The proposed efficient, low dimensional features exploit the power of shallow deep learning models to achieve higher performance with lower computational cost.
In order to show efficacy of the proposed strategy, we have experimented classification of brain tumor grades and achieved the state of the art performance with the resolution of 256 × 256. We also conducted a comprehensive set of experiments to analyze the effect of each component on the
performance.\\



keywords: Brain tumors classification, Medical image classification, Machine Learning, Deep Learning
\end{abstract}

%
%


\section{Introduction}

 
\label{sec:intro}
A primary brain or spinal cord tumor is a tumor that starts in the brain or spinal cord. According to \cite{Siegel2021cancer} an estimated 24,530 adults (13,840 men and 10,690 women) in the United States will be diagnosed with primary cancerous tumors of the brain and spinal cord in 2021. Brain tumors account for 85\% to 90\% of all primary central nervous system (CNS) tumors.
Survival rates depend on several factors, such as age, gender, and type of tumor. Early detection of tumors not only increases the survival chance, but also improves the life quality of patients. Therefore, accurate and noninvasive tumor detection methods are of high interest. Hence, automatic tumor classification methods have gained interest in recent years\cite{kaur2017quantitative, usman2017brain, mohsen2018classification}.\\
Nowadays, most medical image scans contain a sequence of image slices. The process of labeling each slice is time consuming, requires expert knowledge and experience, and is prone to errors. Moreover, due to confidentiality issues of the patients there are lack of publicly available datasets to work with. On the other hand, tumors occur in different locations of the brain and come in various sizes and shapes. Therefore, strong priors cannot be utilized to localize the tumors, and eventually specify their type or grade.
Most of the recent methods\cite{anaraki2019magnetic,sultan2019multi,edraki2020subspace} apply deep learning and use image augmentation to account for the lack of existing labeled data. While these methods show some good results, they usually enforce high computational costs which is not applicable in most hand held devices.
This paper presents a tumor classification system based on features extracted from two domains; Discrete wavelet transform (DWT) and Discrete cosine transform (DCT). The most important aspect of the selected transforms is that they provide informative and sparse features. Next, the features are given into a Long short-term memory (LSTM) Neural Network to account for the sequence nature of the data slices.

The rest of the paper is organized as follows: in section \ref{sec:system Description} a detailed explanation of feature extraction techniques and the
classifier is provided. The proposed approach is described in section \ref{sec:proposed system} followed by the experimental setup and results in section \ref{sec:Experiments}.
Finally, section \ref{sec:Conclusion} concludes the paper.


\section{feature extraction and classifier}
\label{sec:system Description}

We Extended our previous work on brain tumor classification\cite{Fasihi2020Employing} from using a single image (with label per image), to a sequence of images having one label only for the entire sequence. This is the case in many medical image scans present at practice in clinics.
In our method some hand crafted features are extracted
from images. Next, these features are given to an LSTM classifier to precisely categorize the tumor's grade.
There are numerous features that can be used for this purpose such as wavelet, DCT, Haralick texture features\cite{haralick1973textural}, etc. These features serve as coefficients of the basis functions of
the transformed image, where the basis functions are simply
determined by the feature domain.\\ 
Since organ edges and boundaries are not sharp enough in medical images, simple edge detectors are not useful in determining the region of interest (tumor lesion), so this emphasizes on using texture features. Among existing texture features, wavelet transforms have gained a lot of attention in the past few years\cite{mohsen2018classification}. Other than texture features, transforms that summarize information and provide compression are next candidates. DCT is one of the most popular transformations capable of this. Thus combining the abilities of these two transformations together for feature extraction will be of value.
The following two subsections provide more detail about these
two transforms.
\subsection{Discrete Wavelet Transform (DWT)}
\label{ssec:DWT}
DWT keeps both time and frequency
information simultaneously, and is based on multi-resolution analysis (MRA); different frequency bands provide
useful information for image processing. 
Global characteristics are localized in the low frequency band
while the sharp edges of the image are located in high
frequency bands. Since medical images do not contain sharp edges, the majority of information is kept at the approximation image, therefore only the approximation band is kept and the rest is discarded\cite{Fasihi2020Employing}.

\subsection{Discrete Cosine Transform (DCT)}
\label{ssec:DCT}
The DCT represents an image in terms of a sum of sinusoids with different frequencies and amplitudes. This transform concentrates most of the signal power in a small part of the domain, which is proven to be the upper left corner of the transformed image in DCT domain As shown in \cite{Fasihi2020Employing}. Thus, fewer coefficients are sufficient to approximate the original signal, leading to sparse features.
\subsection{classifier selection}
\label{ssec:LSTM}

As mentioned earlier in section \ref{sec:intro}, labels are given for each patient not each image scan individually; Since there exist a sequence of slices for each patient, LSTM network is used for the classification purpose in this study. 
LSTM is a recurrent neural network (RNN) architecture used for classification and regression specially in time series data\cite{hochreiter1997long, graves2012long}. LSTMs were developed to deal with the vanishing gradient problem that can be encountered when training traditional RNNs.
Unlike standard feedforward neural networks, LSTM has feedback connections. These feedback connections are used to store representations of recent input events in form of activations ("short-term memory", as opposed to "long-term memory" embodied by slowly changing weights). Therefore, it can process the entire sequence of images (data) as well as single images.
\begin{figure*}[htb]
  \centering
   \includegraphics[width=0.88\textwidth]{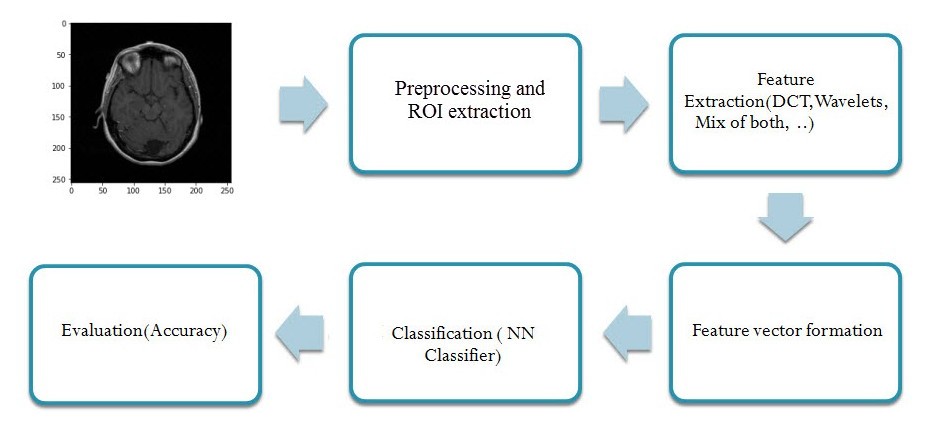}
%
\vspace{-6 mm}
\caption{Block Diagram of proposed classification procedure.}
\label{fig:block_diagram}
\end{figure*}
\section{proposed system}
\label{sec:proposed system}


The block diagram of the proposed classification algorithm is shown in Fig.\ref{fig:block_diagram}. First, all images are preprocessed. Preprocessing step consists of resizing the images to 256x256, and normalizing them to [0, 1] range. Next, K-means clustering is applied on the images to segment the tumor lesion in images. k is selected to be 3 in our experiments. Cluster 1 contains both nontumor and tumor region, while the second cluster gives a shallow boundary of tumor pixels. It is observed that the third cluster contains the region of interest (ROI/tumor). So the third cluster is used. Next, the features are calculated from the ROI, which are then given to LSTM network to classify the tumor grade.
In order to study the effect of both features and LSTM network, we have adopted two baselines. In first baseline the features are given to a multi-layer perceptron (MLP) network, which is a feed forward neural network (Network baseline). Also in order to investigate the effect of transformation and selected features, The ROIs are given directly to the LSTM network without applying feature selection (feature baseline/ second baseline).


\section{Experiments}
\label{sec:Experiments}
\subsection{Data-set}
\label{ssec:data-set}
\begin{figure*}[h]
  \centering
   \includegraphics[width=0.95\textwidth]{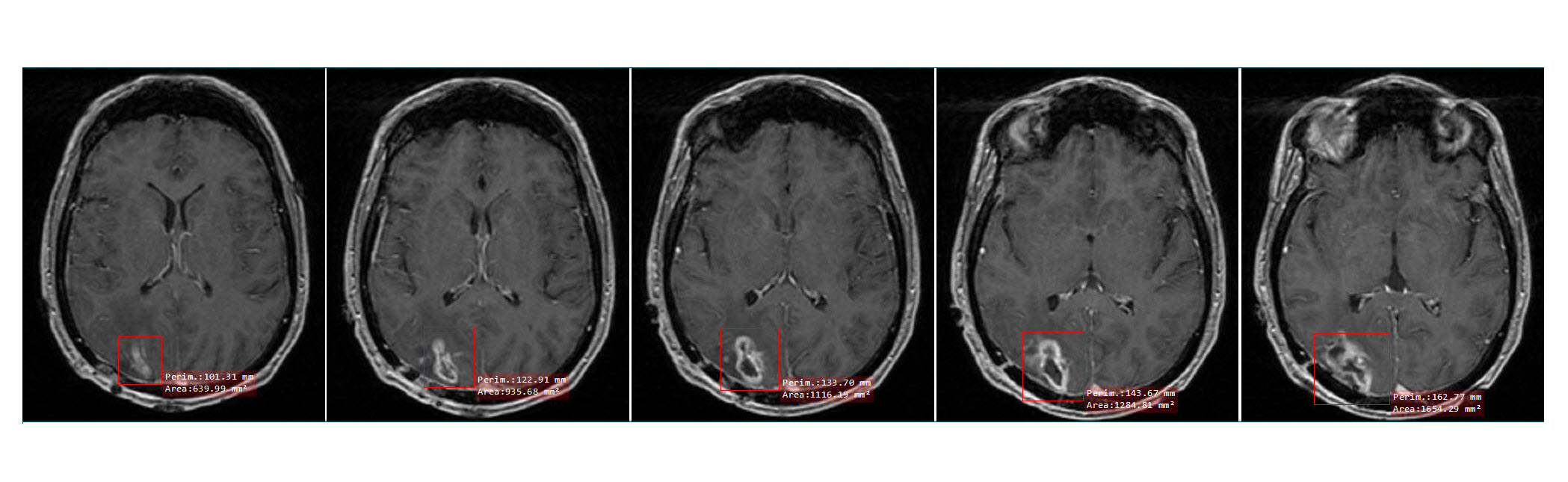}
%
\vspace{-6 mm}
\caption{Sample sequence of dataset for one patient with grade2 glioma Tumor for 5 consecutive slices Axial T1 GD enhanced MRI.}
\label{fig:sample sequence}
\end{figure*}

\begin{figure}[htb]
\begin{minipage}[b]{1.0\linewidth}
  \centering
  \centerline{\includegraphics[width=1\linewidth]{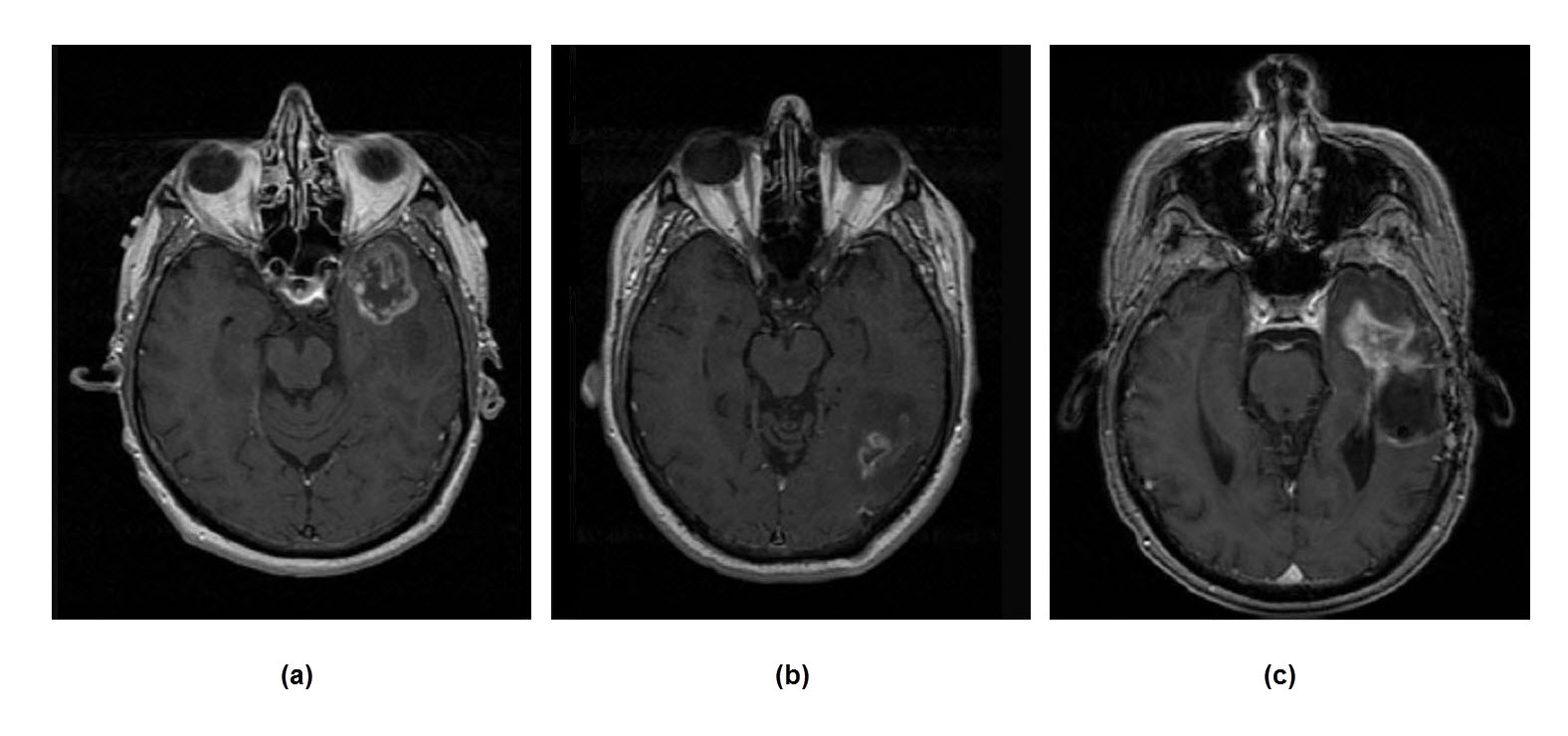}}
\end{minipage}
\hfill
\vspace{-6mm}
\caption{Examples of three different grades of Gliomas T1 GD enhanced axial brain images: (a) Glioma Grade II ; (b) Glioma Grade III; (c) Glioma
Grade IV.}
\label{fig:sample grades}
\end{figure}

To evaluate the proposed method, we used the publicly available brain tumor data set from the Repository of Molecular Brain Neoplasia Data (REMBRANDT) 
\cite{clark2013cancer, scarpace2015data}(obtained from cancer imaging archive). This dataset contains 4404 MRI images from 80 patients (T1 weighted contrast enhanced axial) in total. Each patient is diagnosed with one of the three grades of glioma (Grade II, Grade III, and Grade IV) tumors. 80 patients MRI images with 80 labels (1 label per patient) is used. All images are resized to 256x256 and then the rest of the processing is applied.\\
Fig.\ref{fig:sample sequence} shows a sample sequence of dataset for one patient for 5 consecutive slices (Axial T1 GD enhanced MRI). Also, a sample image for each tumor grade for the same dataset is shown in  Fig.\ref{fig:sample grades}. It is obvious from Fig.\ref{fig:sample sequence} and Fig.\ref{fig:sample grades},  that while there isn't significant interclass variation among the tumors, there exist notable intraclass difference among them; which makes the detection of tumors a challenging task even for experts.

Our experiments consists of two parts: Feature selection, and choosing the appropriate classifier. All experiments were conducted using python Python 3.7. and Keras framework\cite{bloice2016tutorial} is used for our model implementation.
\subsection{Feature selection}
\label{ssec:feature_select}
Features extracted from ROI region are DWT, DCT and mixed transforms of DWT, and DCT proposed in \cite{Fasihi2020Employing}. The mixed features are calculated as follows: first, 2D DWT is applied on each images. Then, 2D Discrete cosine transform is applied only on the LL band (low frequency band) of the transformed wavelet image. The dominant coefficients of the DCT are kept and the rest of the features (residual) are transformed back into the wavelet domain. The wavelet transform is then applied several times to get the DWT feature matrix. The final feature vector is the concatenation of the vectorized DCT and DWT (last LL band) matrices.
In the case of DWT, Haar wavelet is used as the mother wavelet and the general wavelet decomposition structure is utilized when applying the wavelet transform. As mentioned earlier, only approximation level images are used.\\  
For the sake of clarity we define the following notations through out the section.  number of DCT coefficients are shown with ${C_p}$, where ${C}$ refers to DCT and \(p \in \{25, 100\}\). Similarly, number of DWT coefficients are shown with ${W_q}$, where ${W}$ refers to DWT and $q \in \{ 64, 256\}$.
The results of comparing different feature vectors are summarized in Table \ref{tab:accuracy_feature selection} ( Only the best results are reported in the Table ). Based on the results, the combination of transformation improves the classification
accuracy compared to one of the transformations. Also, when the vectorized ROI is used individually (no transformation is applied), 65\% of tumor grades are classified correctly.
\subsection{Classifier}
\label{ssec:classifier}
In the second part, we have utilized dense multilayer neural network classifier as our network baseline. The model has been trained using Adam optimizer \cite{kingma2015adam} with learning rate of $0.005$ with $\beta_{1}=0.9$ and $\beta_{2}=0.999$. 
Rectified linear unit (ReLu) is applied as the activation function on all layers except the last one. To avoid over-fitting on the training data, drop out with ratio of $0.2$ is adopted on the first 2 dense layers.
20\% of the data was used for test and 80\% for training purpose.

\begin{figure}[h]
  \centering
   \includegraphics[width=7cm]{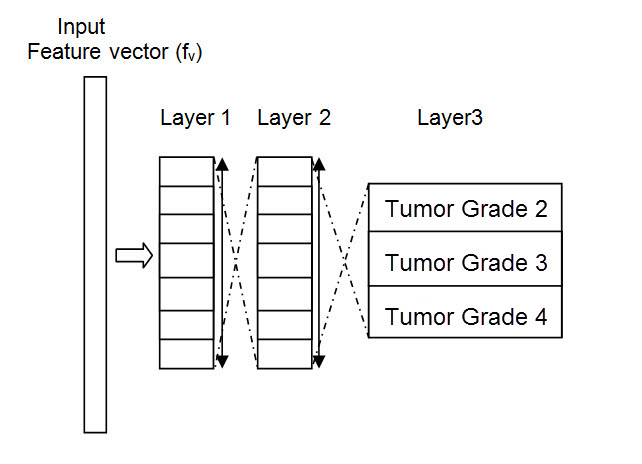}
\caption{Architecture of the NN classifier (baseline network).}
\label{fig:baseline network}
\end{figure}

The 3-layer dense neural network (baseline network) is shown in Fig.\ref{fig:baseline network}. The architecture of this classifier is as follows:
\begin{itemize}
  \item Two Fully connected layers with 150, 90 neurons for first and second layer respectively. Each layer is then followed by dropout and batch normalization 
  \item The last layer, is fully connected layer with 3 neurons, to classify the brain tumor grades.
  \item Number of epochs for training was set to 200, and the learning rate 0.005 is selected with learning rate decay 0.1 after 30 epochs.
\end{itemize}

\begin{figure}[htb]
  \centering
   \includegraphics[width=8cm]{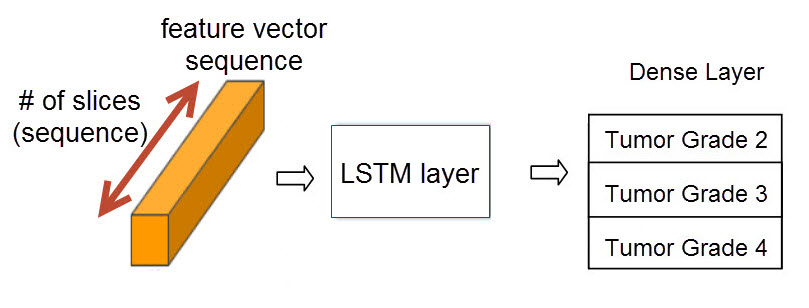}
\caption{Architecture of the LSTM classifier.}
\label{fig:LSTM network}
\end{figure}

For the LSTM network used in this study only one layer of LSTM with output shape of [number of sequence, N] is used. The layer is  followed by a dense layer with 3 neurons, to classify the brain tumor grades. Architecture of this network is shown in Fig.\ref{fig:LSTM network}. Features obtained from the ROI are vectorized and then given to the LSTM layer. Number of sequence is selected to be 30 since it's the minimum number of slices available for each patient in the dataset. N is selected from  \{21, 32\}\ .
The result of comparing the network is summarized in Table\ref{tab:proposed different classifiers}.

\begin{table}[h] 
    \centering
    \renewcommand{\arraystretch}{1.1}
    \caption{ Tumor grade classification Accuracy based on different features (W and C refer to DWT and DCT respectively). Feature baseline refers to when no transformation is applied.}
    \label{tab:accuracy_feature selection}
    \begin{tabular}{|p{2cm}|p{1.3cm}|p{1.7cm}|p{1cm}|p{1cm}|}
       \hline
       {\bf Features} &  {\bf Average Accuracy} & {\bf Best    Accuracy }\\
       \hline 
        $W_{64}$  & 75.02\% & 79.9\% \\
       \hline
       $C_{100}$ & 78.39\% & 80.01\% \\
       \hline
       $C_{100}W_{64}$ &  84.12\% & 86.98\% \\
       \hline
       Feature Baseline  &  65.87\% & 68.32\% \\
      \hline
      \end{tabular}
    
\end{table}

\begin{table}[h] 
    \centering
    \renewcommand{\arraystretch}{1.1}
    \caption{Effect of Changing Classifier Architecture on Accuracy \& Training Time ( with $C_{100}W_{64}$ features ). }
    \label{tab:proposed different classifiers}
    
    \begin{tabular}{|p{1.5cm}|p{1.5cm}|p{1.3cm}|p{1.2cm}|p{1cm}|}
       \hline
       {\bf Architecture} & {\bf {parameters} }&  {\bf Average Accuracy} & {\bf Best \mbox{Accuracy} }&  {\bf Training Time(sec)}\\
       \hline
        LSTM Network 21   & 7290 &  84.12\% & 86.98\% & 570.022 \\
        \hline
        LSTM Network 32  & 12515  &  81.5\% & 84.41\% & 616.167\\
        \hline
        Baseline Network   & 302973 &  75.02\% & 78.98\% & 261.068 \\
       \hline
      \end{tabular}
    
\end{table}

The number of parameters of LSTM classifier is about 40 times less than the baseline Network, but the  processing time is around 2.2 times the baseline network. This is due to complex nature of LSTM layer compared to feed-forward NN. This increase of time can be traded off with the increase in performance.


According to Table \ref{tab:accuracy_feature selection}, the best result was obtained when C100 and W64 feature vector along with LSTM architecture 21 (LSTM with output shape of [number of sequence=30, 21]), was used. The DCT part in the feature vector provides the important features of the image, while the DWT part of the feature vector provides texture features not captured by DCT.

\section{Conclusion}
\label{sec:Conclusion}
In this paper, a brain tumor grade classification system is proposed. The system is based on utilizing an optimal and compressed set of features from DWT and DCT domains. This set is then fed into LSTM classifier. Various options were investigated both in feature selection and classifier architecture. The results of the study indicates that the proposed system that is using the combination of DWT and DCT increases the accuracy without significant time overhead  compared to employing each of the transforms individually. Moreover, the proposed system outperforms the baseline network in terms of accuracy.
Overall, the proposed method is generally applicable on many existing medical image scans, where not every single image is labeled.  On the other hand, the technique presented here can be implemented on handheld devices for medical applications. 


\bibliographystyle{ieee}
\bibliography{refs}




\end{document}